\documentclass[acmsmall,screen,nonacm]{acmart}

\setcopyright{none}
\copyrightyear{2018}
\acmYear{2018}
\acmDOI{10.1145/1122445.1122456}

\acmPrice{15.00}
\acmISBN{978-1-4503-XXXX-X/18/06}

\usepackage{pifont}
\newcommand{\yes}{\ding{51}} 
\newcommand{\no}{\ding{55}} 

\usepackage{extdash}

\usepackage{csquotes}

\usepackage{listings}

\usepackage{lstlangcoq}
\newcommand{\langCoq}{\lstset{language=Coq,basicstyle=\ttfamily\small,columns=fixed,basewidth={0.5em}}}
\newcommand{\langHaskell}{\lstset{language=Haskell,basicstyle=\ttfamily\small,columns=fullflexible,commentstyle=\color{gray},deletekeywords={Monad,Integral,Int,Bool,Maybe,fromIntegral}}}
\langHaskell

\newcommand{\riscv}{RISC\nobreakdash-V} 
\newcommand{\sail}{Sail} 

\newcommand{\instance}[2]{\noindent\fbox{\parbox{\textwidth}{Support \textbf{#1} by #2.}}}

\begin{document}

\title{Flexible Instruction-Set Semantics via Type Classes}

\renewcommand{\footnotetextcopyrightpermission}[1]{}

\newcommand{\socialdistancing}{, }

\author{{Thomas Bourgeat\socialdistancing Ian Clester\socialdistancing Andres Erbsen\socialdistancing Samuel Gruetter\socialdistancing Pratap Singh\socialdistancing Andrew Wright\socialdistancing Adam Chlipala}
} 
\affiliation{
  \institution{MIT CSAIL}
  \country{USA}
}

\makeatletter
\let\@authorsaddresses\@empty
\makeatother


\begin{abstract}
  Instruction sets, from families like x86 and ARM, are at the center of many ambitious formal-methods projects.
  Many verification, synthesis, programming, and debugging tools rely on formal semantics of instruction sets, but different tools can use semantics in rather different ways.
  As a result, a central challenge for that community is how semantics should be written and what techniques should be used to connect them to new use cases.
  The best-known work applying single semantics across quite-different tools relies on domain-specific languages like Sail, where the language and its translation tools are specialized to the realm of instruction sets.
  We decided to explore a different approach, with semantics written in a carefully chosen subset of Haskell.
  This style does not depend on any new language translators, relying instead on parameterization of semantics over type-class instances.
  As a result, a semantics can be a \emph{first-class object within a logic}, and application of a semantics for a new kind of tool can be \emph{a first-class operation in the logic}, allowing sharing of theorems across applications.
  Our case study is for the open RISC-V instruction-set family, and we have used a single core semantics to support testing, interactive proof, and model checking of both software and hardware.
  We especially highlight an application of a first-class semantics within Coq that can be instantiated in different ways within one proof: simulation between variants where multiplication is implemented in hardware or in the machine code of a particular software trap handler.
\end{abstract}

\maketitle

\section{Introduction}

Machine-language instruction sets are at the center of many aspects of systems implementation and verification.
Such an important interface deserves a formal semantics.
One semantics should be usable for all of documentation, simulation, testing, model checking, and interactive theorem proving.
Furthermore, it should be usable for all the above applied to both software and hardware.

Many past projects demonstrated individual semantics usable for minorities of these cases.
Leading approaches like \sail~\cite{SAIL_POPL19} involve domain-specific languages (DSLs) and ad-hoc translators from them into different languages appropriate to different use cases.
Certainly, in many ways, it is hard to compete with languages purpose-built for a given style of program.
It is not hard to stock a bestiary of potential domain-specific specification languages: high-level-language semantics, distributed-systems invariants, temporal logic for hardware, even pure mathematics.
All the above might be brought together in one end-to-end-verified system for e.g. a cryptographic protocol implementation, with different layers verified using different tools and styles.
However, there is a lurking risk of a classic ``$n^2$'' compilers problem: a translator must be built between each relevant pair of specification language and input language of a formal-methods tool.
We might also worry about learnability challenges, for when one engineer encounters a specification in a new domain and must get to know a new language and not just a new library.

The two extremes of fully domain-specific and fully general-purpose specification languages present a challenging space of trade-offs, and it will take much more work to approach anything like a universal answer on which is ``better.''
However, in the study we report in this paper, our goal was to demonstrate that an important specification category, of instruction sets, can be handled pragmatically within a general-purpose specification language.
Applications of our semantics for \riscv{} achieved the following firsts for that instruction-set-architecture (ISA) family, to our knowledge.
\begin{itemize}
\item The first use of one semantics for both model-checking and interactive proof of deep functional-correctness properties of \riscv{} machine code
\item The first use of one semantics for both model-checking and interactive proof of deep functional-correctness properties of \riscv{} processor implementations
\end{itemize}
Further, our semantics was at the center of the work of \citet{lightbulb}, one of only a few results so far of deep functional-correctness proof of both hardware and complete software applications, formally linking their theorems through the common interface of an instruction-set semantics.
Each use case involves different choices of complexities to bring in (e.g., virtual memory? weak memory model? I/O device interface?).
We also support direct execution of the semantics with decent performance, and the original form of the semantics is decently readable (if not yet writable) even by an audience with the standard backgrounds of software or hardware engineers.
It is difficult to be sure how much more or less smoothly these different exercises would have gone using other specification styles.
Rather than trying to assert a clear directional comparison, we only aim to show that a promisingly complete ecosystem of tools can be stood up at reasonable cost, around a semantics written in a general-purpose language.

Our prototype semantics is implemented in a relatively small subset of Haskell.
An advantage of choosing such a (relatively!) popular language is that translators for it already exist to a variety of other relevant languages; our case studies use translators to Coq and Verilog.
The specific language and set of translators are somewhat incidental to our larger message.
The right common specification language of the future might very well be quite different, but at least it would still only need one translator per target language, avoiding the $n^2$ problem we warned about above.
However, one celebrated Haskell feature is central to how we make a single specification very flexible.

\citet{Monads_POPL92} introduced monads in functional programming as a way to write code that is abstracted over kinds of effects.
Indeed, they had already been used in projects like the Lava hardware framework~\cite{Lava} to allow flexible alternative interpretations of definitions.
We will show that essentially the same kind of abstraction is a good fit for the variation across uses of instruction-set specifications.
All the work of adapting the semantics for a new use case is in defining an instance of our new type class that extends monads with operations relevant to \riscv{} execution.
In other words, reusing our semantics is more like the standard experience of using a library than of picking and choosing generators associated with a DSL.
An advantage of library status is that the semantics has first-class status, so, for instance, mechanized proofs of important metatheorems can be completed once-and-for-all.
We will demonstrate this last point with a Coq proof that relates two instantiations of the semantics, one where the multiply instruction is implemented in hardware and one where it is implemented in software, by a trap handler for unimplemented instructions.
The two instantiations are merely choosing different monads (three, in fact) to run a single semantics with.
Instantiation takes place within the logic, rather than via tooling that generates new source files with no a-priori connection across instantiations.

Let us give a bit of background on \riscv{}, the instruction set we chose, before turning to detail on how the semantics is written and how it is applied.

\subsection{RISC-V as a formal-methods-research enabler}

Until recently, formal-methods projects requiring specifications of processors were in an uncomfortable situation:
the ISAs used in real processors were very complex, did not have openly accessible specifications, or were protected by patents.
Therefore, each formal-methods project either invented its own ISA (e.g. \citet{CakeMLSilver_PLDI19, bedrock_ICFP13}) or formalized from-scratch, at the desired abstraction level, a small subset of an existing ISA (e.g. \citet{Leroy-Compcert-CACM}).

Basic researchers breathed a sigh of relief with the increasing popularity of \riscv{}\footnote{\url{https://riscv.org/}}, which is controlled by a nonprofit foundation that does not limit who may implement the architecture.
It is particularly compelling for formal-methods research, because it is a simple clean-slate design easy to reason about and yet features a mature open ecosystem.
\riscv{} is a family of instruction sets broken into different native bitwidths (32, 64, 128) and extensions (e.g., multiplication and division instructions, atomics, \dots) that may be mixed and matched.
On the software side, there are fully upstreamed versions of Clang, GCC, and Linux.
On the hardware side, the set of commercial and open-source processor designs is growing rapidly~\cite{sifive, westerndigital, xuantie, kendryte, rocket, boom, riscyooo, zeroriscy}.
It makes for a very compelling platform for experimentation in basic research, progressively extensible to full-fledged realistic systems, where it is easy to tinker with any part of the software-hardware ecosystem.

Another advantage of \riscv{} is existing community momentum around formal specification.
There is an institutionally blessed semantics in the Sail DSL.
While we would like to prove our semantics against that one eventually, we are held back for the moment by precisely the $n^2$ problem we described earlier, as we found the Sail-to-Coq translator to output unusably verbose code that was very challenging to work with interactively.
That behavior is not surprising for a relatively young Sail tool mostly sharing code oriented around fast simulation via C code, and this weakness might very well be corrected soon, at which point we would be glad to undertake this exercise (no longer needing to do significant Sail-specific tool hacking ourselves).

\subsection{Outline}

The remaining sections review important elements of the \riscv{} ISA, explain our specification style, and discuss how to cover different use cases with different type-class instances.
Our implementation, including case studies, is available as a supplement to this submission.

The core claim we aim to defend throughout is that \textbf{a formal semantics for an industrial-strength instruction-set family can be implemented fruitfully in a general-purpose language and applied across a representative set of examples, in formal methods and elsewhere, without writing any new spec-language translators}.
We defend that claim by showing how each example works merely by choosing the right type-class instances and perhaps also by calling a from-Haskell translator that already existed.
We also aim to illustrate \textbf{the benefits of first-class parameterized semantics}, where a semantics can be specialized to new applications within a proof assistant, allowing proof of metatheorems that apply to all specializations.

\section{Overview}\label{sec:overview}

\begin{figure*}
  \begin{center}
    \includegraphics[width=\linewidth]{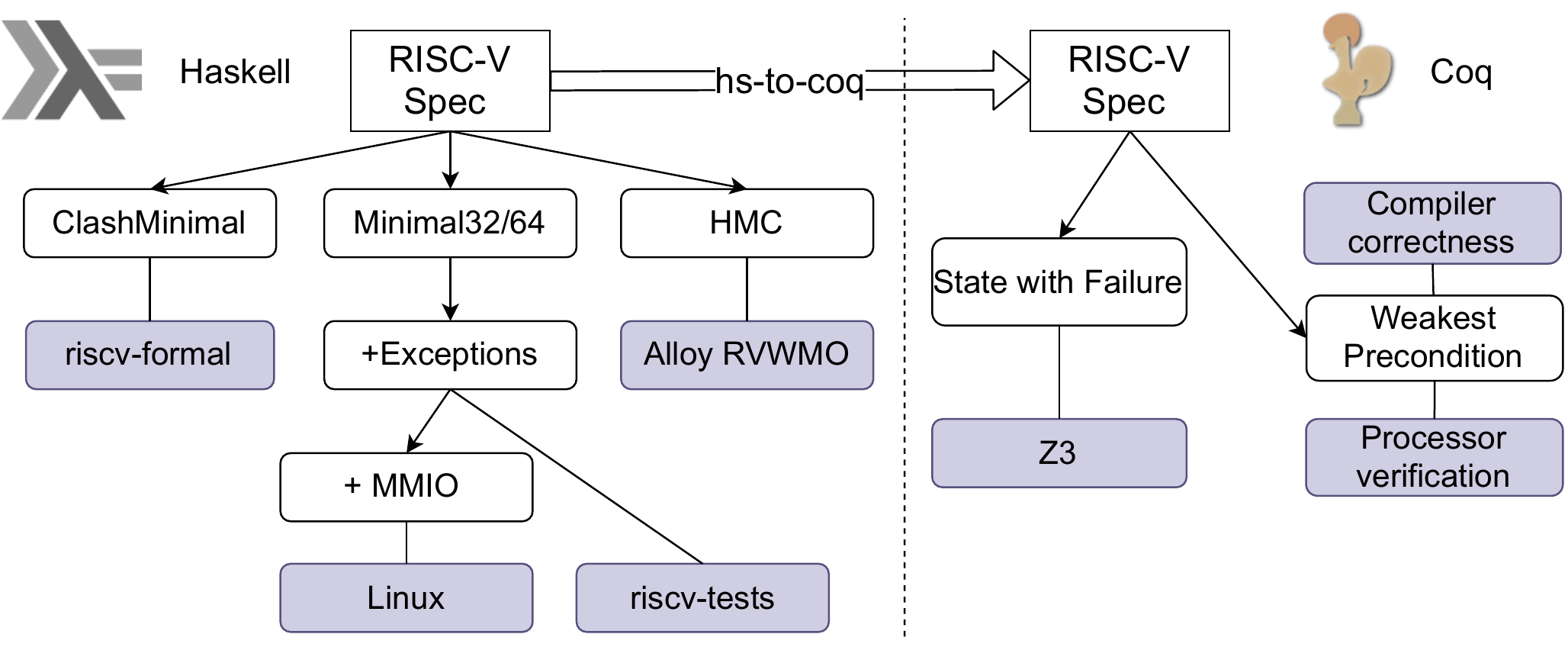}
  \end{center}
        \caption{Projects using our \riscv{} specification. The boxes with rounded corners are type-class instantiations. Instances with names that start with ``+'' are derived from other instances (adding new features). The grey boxes show external projects that our specification interacts with.}\label{fig:connections}
\end{figure*}

The nonprofit \riscv{} Foundation (since renamed \riscv{} International) developed an instruction-set-architecture specification in English~\cite{RISCV_User, RISCV_Priv}.
The goal of our project is to translate the English specification into a broadly applicable, machine- and human-readable formal specification.

There is a particular emphasis on avoiding overspecification:
Platform-specific details that are left unspecified by the English specification should also be unspecified in the formal specification, while at the same time enabling users of the formal specification to pin down as many of these platform-specific details as appropriate for their use cases.

Orthogonally, we want to support the many different use cases shown in \autoref{fig:connections}:
For instance, some users want an executable specification that returns one single final machine state, while others might want to leave some inputs, parameters, or nondeterministic choices abstract and obtain a logic formula restricting the set of possible final states, or obtain a list of final states, or obtain a checker that simply answers whether a given execution trace is allowed by the specification, etc.

To summarize, these requirements lead us to the following dimensions of parameterization:
\begin{itemize}
\item Supported extensions (see \autoref{tab:extensions}) and bitwidth (32\Hyphdash bit, 64\Hyphdash bit or left abstract)\footnote{The \riscv{} specification also defines a 128-bit variant that we did not consider.}
\item Platform-specific details
\item Use-case-specific details
\end{itemize}

\subsection{Choice of language}

Serious commitment to a multipurpose specification requires careful thought about the language it should be written in.
One goal was to write readable functional programs that could be understood \emph{intuitively} by hardware engineers or compiler hackers, even if they are not familiar with the underlying features (such as monads, type classes, etc.) enabling readability and parameterizability of the specification.
Further goals were to create a specification that is practical to use in interactive theorem provers and to connect to other specifications, especially from the hardware world.

We found that Haskell was able to cover most of the constraints, thanks to the following:
\begin{itemize}
        \item \lstinline{do} notation provides syntactic sugar for readable imperative\Hyphdash looking code, particularly useful for this specification.
        \item the Clash compiler~\cite{clash}, compiling Haskell (with bounded recursion and finite datatypes) to Verilog/VHDL, is a bridge to hardware-model-checking tools.
        \item hs-to-coq~\cite{hstocoq_ICFP18}, a compiler that uses the GHC frontend to generate Haskell-like Coq code, built to prove Haskell programs in Coq, is a good bridge to interactive theorem proving in Coq.
\end{itemize}

We restrict ourselves to concepts supported by these three Haskell compilers (hs-to-coq, Clash, GHC).
Via hs-to-coq, we produced a semantics that was chosen as the reference machine model in several Coq projects.
Via Clash, we produced a minimal ``single-cycle'' Verilog execution model for which external people (authors of other specifications) checked the agreement between our spec and theirs.
Finally, via GHC, we demonstrated the possibility to explore the basics of the \riscv{} memory model and to test our specification as a simulator.

In our work, Haskell is a convenient stand-in for whatever turns out to be the ideal general-purpose language for writing a variety of multipurpose specifications.
Writing translators to the input formats of different formal-methods tools can be a significant undertaking, so that it pays to introduce as few languages as possible that require translation.
It may turn out that the ideal community-shared spec language is quite different from Haskell, but our interest here is showing that a general-purpose language can work well as the host for a multipurpose spec, leaving fine-tuning of the approach for future work.
An important point is that we avoid use of any spec-language features specialized to ISAs.

\subsection{Structure of the specification}

Our \riscv{} specification is composed of several \emph{extensions} listed in \autoref{tab:extensions}, and an implementation can choose which subset of them to support.
Our formalization of the specification only covers the most important of them.

\begin{table}
\begin{tabular}{lcccc}
\toprule
        Description                   & Name     & hs?  & Coq?  & Clash?\\
\midrule
        Integer                       & I        & \yes & \yes & \yes \\
        Integer Multiply/Divide       & M        & \yes & \yes & \no \\
        Atomics                       & A        & \yes & \no & \no \\
        Single Floating-Point         & F        & \yes & \no & \no  \\
        Control \& Status Registers   & Zicsr    & \yes & (\yes) & \no \\
\bottomrule
\end{tabular}
\caption{Standard Extensions of \riscv{} 20191213 (excluded: E, D, Q, L, C, B, J, T, P V, N, Zifencei, Zam, Ztso)}\label{tab:extensions}
\end{table}

\paragraph{The primitives}

The key to supporting many different use cases is to specify the semantics of each instruction in terms of a small number of \emph{primitives} listed in \autoref{code:primitives}, while leaving the implementations of these primitives to be filled in by the concrete use cases.
The primitives include state-like constructs (for the registers, the memory, \dots) plus control-flow-like constructs (\lstinline{endCycle}) to capture the control-flow change in case of an exception (see \autoref{exceptions}) raised in the middle of the semantics of a function (an early return).

\begin{figure}\small
\begin{lstlisting}
-- Indicates which stage is the source of a memory access
data SourceType = VirtualMemory | Fetch | Execute
-- Type class providing the RISC-V primitives:
class (Monad p, MachineWidth t) => RiscvMachine p t | p -> t where
  getRegister :: Register -> p t
  setRegister :: Register -> t -> p ()
  getFPRegister :: FPRegister -> p Int32
  setFPRegister :: FPRegister -> Int32 -> p ()
  loadByte :: SourceType -> t -> p Int8
  loadHalf :: SourceType -> t -> p Int16
  loadWord :: SourceType -> t -> p Int32
  loadDouble :: SourceType -> t -> p Int64
  storeByte :: SourceType -> t -> Int8 -> p ()
  storeHalf :: SourceType -> t -> Int16 -> p ()
  storeWord :: SourceType -> t -> Int32 -> p ()
  storeDouble :: SourceType -> t -> Int64 -> p ()
  makeReservation :: t -> p ()
  checkReservation :: t -> p Bool
  clearReservation :: t -> p ()
  getCSRField :: CSRField -> p MachineInt
  unsafeSetCSRField :: (Integral s) => CSRField -> s -> p ()
  getPC :: p t
  setPC :: t -> p ()
  getPrivMode :: p PrivMode
  setPrivMode :: PrivMode -> p ()
  commit :: p ()
  endCycle :: forall z. p z
  flushTLB :: p ()
  fence :: MachineInt -> MachineInt -> p ()
  getPlatform :: p Platform
\end{lstlisting}
\caption{The primitives of the abstract RISC-V monad}\label{code:primitives}
\end{figure}

\paragraph{Missing axioms}
One use of our semantics is verification of properties that hold for multiple (even infinitely many) different possible instantiations of this type class.
In that way, we are able to establish \emph{metatheorems} that hold across applications of the semantics, a goal hard to achieve when new applications are supported with new code produced by ad-hoc translators.
However, for now, we do not support proving deep properties of our semantics for any possible instance of this type class, rather imposing additional restrictions on instances where needed.
The reason is that we are not using \emph{axiomatic type classes}~\cite{axiomatic_type_classes} that additionally assert logical axioms that any instance must validate.
For instance, one might hope that, for any valid instance, getting the value of a register that we just wrote should return the value we have just written.

The trouble is that what counts as ``reasonable'' becomes much murkier very quickly, as we consider other aspects of the semantics.
Foundational research questions remain around e.g. memory models in the presence of multisize accesses, virtual memory, self-modifying code, etc.
These questions may come to be sufficiently settled to allow a complete-enough set of axioms to be included in our type class.
For now, though, we explore just (1) the benefits of allowing the code for the instruction definitions themselves to be shared fruitfully across a variety of tools and (2) the possibilities of instance-parameterized theorems within Coq, with further local constraints.

\paragraph{Instantiation}
The abstract monad \lstinline{p} (of kind \lstinline{* -> *}) can be instantiated differently by each use case, which keeps our spec agnostic to the concrete state of the machine and to the kinds of effects that instructions can have.
For instance, depending on the platform and the use case, an invocation of the \lstinline{storeWord} primitive could update the memory of the machine state, or it could fail if the address is outside of the physical address range, or it could record constraints in a memory-model graph, or it could record an I/O event if the address is in a range that the platform uses for memory-mapped I/O, etc., and our specification is completely agnostic to these options.

The abstract type \lstinline{t} is the type of the values stored in the integer registers.
It can be instantiated with \lstinline{Int32}, \lstinline{Int64}, or left abstract for use cases where it makes sense to reason about all bitwidths at once.
Requiring a \lstinline{MachineWidth} type-class instance for \lstinline{t} guarantees that there are arithmetic and logical operators for \lstinline{t}.
To distinguish \lstinline{t} from helper integer values that do not live in registers, we introduce an additional integer type \lstinline{MachineInt}, which is an alias for \lstinline{Int64}, and whose more-significant bits are sometimes ignored.
In fact, whenever we need an $n$-bit integer (with $n \leq 64$) that does not live in a register, we use \lstinline{MachineInt}, applying bitmasking where necessary.
In this way, we dodge use of dependent bitvector types indexed by width, as provided by DSLs like Sail.
The tradeoffs on this question may also be debated, but at least representation is eased in general-purpose languages with rich but not dependent types, and our case studies will demonstrate that a variety of tasks in programming and formal methods remain tractable with laxer help from typing.

\paragraph{Instructions}
The above choice also shows up in our algebraic datatype of instructions:
\begin{lstlisting}[basicstyle=\ttfamily\small]
data InstructionI =
  Sw { rs1 :: Register, rs2 :: Register, simm12 :: MachineInt } |
  Add { rd :: Register, rs1 :: Register, rs2 :: Register } |
  Beq { rs1 :: Register, rs2 :: Register, sbimm12 :: MachineInt } | ...
\end{lstlisting}

For instance, even though the offset field of the store-word instruction is only a signed 12-bit immediate, we represent it with a (64-bit) \lstinline{MachineInt} for simplicity.
Note that this simplification does not compromise correctness, because the specification only creates instructions in the decoder, which only ever writes 12-bit values into that field.

\paragraph{Decode}
The decoder starts by defining symbolic names for notable bitfields of the instruction \lstinline{inst} being decoded:
\begin{lstlisting}[basicstyle=\ttfamily\small]
opcode  = bitSlice inst 0 7
rd      = bitSlice inst 7 12
rs1     = bitSlice inst 15 20
rs2     = bitSlice inst 20 25
simm12  = signExtend 12 $
  shift (bitSlice inst 25 32) 5 .|. bitSlice inst 7 12
...
\end{lstlisting}
and then defines a decoder for each \riscv{} extension:
\begin{lstlisting}[basicstyle=\ttfamily\small]
decodeI
  | opcode==opcode_STORE, funct3==funct3_SW = Sw rs1 rs2 simm12
  | opcode==opcode_BRANCH, funct3==funct3_BEQ = Beq rs1 rs2 sbimm12 ...
\end{lstlisting}
and finally, checks if the decoded instruction is part of the supported extensions.


\paragraph{Execute}
For each \riscv{} extension supported, there is an \lstinline{execute} function that expresses the effects of each instruction of the extension in terms of the primitives listed in \autoref{code:primitives}.
For instance, here is the definition of the jump-and-link-register instruction, for which explanatory prose will be provided shortly thereafter:

\noindent\begin{minipage}{\linewidth}
\begin{lstlisting}[basicstyle=\ttfamily\small]
execute (Jalr rd rs1 oimm12) = do
  x <- getRegister rs1
  pc <- getPC
  let newPC = (x + fromImm oimm12) .&. (complement 1)
  if (remu newPC 4 /= 0)
    then raiseExceptionWithInfo 0 0 (fromIntegral newPC)
    else (do
      setRegister rd (pc + 4)
      setPC newPC)
\end{lstlisting}
\end{minipage}

It was transcribed from the following English definition:

\begin{displayquote}
The indirect jump instruction JALR (jump and link register) uses the I-type encoding. The target
address is obtained by adding the sign\Hyphdash extended 12-bit I-immediate to the register $rs1$, then setting
the least-significant bit of the result to zero. The address of the instruction following the jump
($pc+4$) is written to register $rd$. Register \lstinline{x0} can be used as the destination if the result is not
required.

The JAL and JALR instructions will generate an instruction-address-misaligned exception if the
target address is not aligned to a four-byte boundary.
\end{displayquote}

It uses Haskell's \lstinline{do} notation to chain monadic operations, and it can also use standard Haskell constructs such as \lstinline{let} or \lstinline{if}.
The binary operators (such as \lstinline{+}, \lstinline{/=} and \lstinline{.&.} in this example) are provided through the \lstinline{MachineWidth} type class.
It inherits from Haskell's standard type classes \lstinline{Integral} and \lstinline{Bits}, allowing us to use the standard infix operators.
\lstinline{MachineWidth} also provides \lstinline{fromImm} to convert \emph{immediates} from instructions into register values \lstinline{t}.

\paragraph{Run}
Finally, we define what it means to run one instruction.
For the Coq proofs, we use a simplified\footnote{but still accurate with respect to real software and hardware, because both the processor and the compiler are proven to respect this same simplified specification} version that just fetches an instruction, decodes and executes it, and updates the program counter, whereas the Haskell version also considers interrupts and exceptions.

\paragraph{Use-case- and platform-specific code}
The components described so far form the specification and are grouped together in a directory called \lstinline{Spec}.
However, we have not yet defined how state is represented and how the primitives of \autoref{code:primitives} are to be implemented.
These use-case- and platform-specific definitions are in a separate directory called \lstinline{Platform} and are the subject of the next section.
The \lstinline{Spec} and \lstinline{Platform} directories each are about three thousand lines of code.

\section{Different monads for different use cases}\label{sec:usecases}

Our \riscv{} specification benefits from using a monad (the \lstinline{p} in \autoref{code:primitives}) in the very same way as Wadler's interpreter in his classic ``The Essence of Functional Programming'' paper~\cite{Monads_POPL92}.
In this section, we tour the wide variety of applications that can be connected to this single semantics.
The key claim we emphasize throughout is that \textbf{each application is supported simply by picking new type-class instance(s)} and perhaps also by calling general-purpose translators from Haskell -- no new translators or even parsers specific to ISA descriptions were required.

We begin with a brief description of mundane use cases in interpretation (essentially the original application of monads that concerned Wadler).
Next, we spend the bulk of our discussion on different flavors of interactive proofs with Coq, before wrapping up with case studies of automated model-checking (outside Coq) of both machine-code programs and \riscv{} processors.

\subsection{Simulation}\label{sec:simulation}

\instance{fast interpretation of machine-code programs}{choosing efficient but harder-to-reason-about data structures}

The most common way of modeling processors in the
formal-methods world is to consider the machine to be deterministic, each cycle
updating the state of the registers and the memory depending on the instruction
present in memory at the location pointed to by the program counter.

Concretely, we wrote two instances of the \lstinline{RiscvMachine} type class, named \lstinline{Minimal32} and \lstinline{Minimal64}, to obtain a 32- and a 64-bit \emph{machine simulator}.
We instantiated the type class in the I/O
monad, using references and arrays to implement registers, program counter, and
memory.






\subsection{Supporting RISC-V exceptions}\label{exceptions}

\instance{early exit via exceptions}{using a variant of the classic monad transformer for failure}

Many formal-methods-oriented projects do not want to deal with
exceptions or interrupts, while others are interested in modeling and
leveraging them.
The formal-spec user should experience the costs of those further features incrementally as they are brought into play.

Raising an exception involves two components: modifying a bunch of state (namely numerous special state registers [CSRs])
and bailing out of a cycle early, not performing the effects of the instruction that would have come
after an exception is raised.
In \riscv{}, exceptions can be caused by virtual-memory translation failures, failed system privilege checks, and alignment problems, among other causes.

To be able to write monadic values carrying this early-exit information, we encode the early return in a layer of a \texttt{MaybeT} monad transformer.
The crucial primitive already appeared in our definition of the \texttt{RiscvMachine} monad.
\begin{lstlisting}[basicstyle=\ttfamily\small]
      endCycle :: forall z. p z
\end{lstlisting}

Conspicuously, an implementation of \texttt{endCycle} is missing from the base machine previously described as \texttt{Minimal64}/\texttt{32}, but there is one given in source file \texttt{Machine.hs}:

\begin{lstlisting}[basicstyle=\ttfamily\small]
instance (RiscvMachine p t) => RiscvMachine (MaybeT p) t where
  getRegister r = lift (getRegister r)
  setRegister r v = lift (setRegister r v)
  ...
  endCycle = MaybeT (return Nothing) ...
\end{lstlisting}

This instance demonstrates something powerful about our spec: composability.
It takes some existing instance of the same type class and builds on it,
adding in the functionality of the \texttt{Maybe} monad.
In that monad, a computation halts as soon as
a step returns \texttt{Nothing}, precisely capturing the ``early-return''
behavior we want upon encountering an exception.


With this two-layer specification, we ran in simulation the \texttt{riscv-tests} test suite (\texttt{rv64mi}, \texttt{rv64si}, \texttt{rv64ui}, \texttt{rv64ua}), which is the
standard community-maintained test suite.

\subsection{Platform modeling, MMIO, and devices}

\instance{I/O features}{providing additional monad transformers}

We actually use this kind of instance augmentation repeatedly in our code, first to encode the
semantics of core features (like \riscv{} exceptions, as we have just seen)
but also to add features like memory-mapped I/O devices to existing
\lstinline{RiscvMachine} instances.

For example, we enrich the platform with a concurrently memory-mapped device: a
UART connected to from a terminal, which generates interrupts
received by the main loop of the simulator.

With this implementation, composed of three layers of specifications, we were able
to run Linux in January 2019  at about 100k instructions per second. (We have not
tracked newer Linux versions.)

While this strategy allows us to experiment, test, and vouch for our good coverage of
the spec, this artifact is not especially competitive for running significant \riscv{} programs,
e.g. the popular QEMU runs at several hundred million instructions per second.

\subsection{Interactive theorem proving}

\subsubsection{Translation from Haskell}

Using hs-to-coq~\cite{hstocoq_ICFP18}, we can translate the Haskell specification to Coq.
Since hs-to-coq was designed to model Haskell semantics in Coq as faithfully as possible, it ships with handwritten and auto-generated translations of Haskell's standard-library files, and by default they are referenced by the Coq files produced by hs-to-coq.
However, for this project, we were not seeking a faithful reproduction of Haskell semantics in Coq but rather an idiomatic \riscv{} specification in Coq.
Therefore, we used hs-to-coq's \emph{edit files} feature, which allows one to provide renaming and rewriting patterns to be applied during the translation, so that we could map all Haskell standard-library references to reasonably close Coq equivalents and obtain an idiomatic, Haskell-independent Coq specification.

We used hs-to-coq to translate the files specifying how each instruction is executed, the instruction decoder, as well as the CSR-file specification, while we manually wrote remaining files like utility definitions, the definition of the \texttt{RiscvMachine} type class, and proof-specific files.

\subsubsection{Use as the interface between software and hardware}

Our \riscv{} Coq specification was used successfully in a project~\cite{lightbulb} that combines a compiler-correctness proof with a processor-correctness proof.

The combined theorem states that the I/O trace produced by the processor matches the one produced by the source program fed to the compiler, without referencing the \riscv{} specification any more.
Thus, auditors of the system can know the behavior of the system without having to audit whether both the compiler and the processor interpret the \riscv{} specification in the same way, which greatly reduces the auditing burden.

\subsubsection{Opting out of features and opting back in}

\quad

\instance{different applications referring to different complexities of the ISA family}{connecting those complexities to methods of the type class}

Our first version of the translation to Coq was driven by the requirements of the compiler-correctness project mentioned above, which required a very simple and manageable spec to get started, so it was decided that initially, CSRs should not be modeled.
However, this also meant that we could not use the real \lstinline{raiseException} function, nor the \lstinline{translate} function (translating virtual to physical addresses), which starts by reading a CSR that indicates whether virtual memory is enabled.
The solution was surprisingly simple:
Since we had already chosen manual translation of the file containing the declaration of the \lstinline{RiscvMachine} type class, we were free to abstract over \lstinline{raiseException} and \lstinline{translate} by adding them to the primitives of \lstinline{RiscvMachine} (\autoref{code:primitives}).
That is, we made our specification \emph{more configurable} than \riscv{} allows, and for the compiler, we instantiated \lstinline{translate} to the identity function and \lstinline{raiseException} to hard failure, because no instructions emitted by the compiler rely on exceptions, while at the same time, we kept open the possibility to instantiate these two functions with (more) real definitions.

These modifications are not \riscv{}-compliant, but we consider it an important feature of our specification that we were able to make them, while still being able to translate most of the specification to Coq automatically and thus continuously pull updates and bugfixes made in Haskell into the Coq code base.

Later, when we added CSRs to the Coq specification, we wrote a simplified \lstinline{raiseException} function.
Since the compiler does not use it, it was trivial to integrate this update with the proof.

\subsubsection{Simulator in Coq}

\quad

\instance{interpretation within a proof assistant}{developing similar instantiations to the earlier simulator case, using more naive purely functional data structures.  A failure monad provides convenient opportunities to indicate unsupported language features, such that individual programs must then be proved not to exercise those features}

\langCoq
\paragraph{State monad}
In Coq, the simplest-possible instantiation of the monad is
\lstinline{p := State MachineState},
where \lstinline{State} is the state monad
defined as
\lstinline{State(S A: Type) := S -> (A * S)},
and \lstinline{MachineState} is a record containing the values of the processor's registers, the program counter, the memory, the CSR file, and the current privilege level.
This instantiation can be used to obtain a deterministic \riscv{} simulator.

\paragraph{State monad with failure}
An arguably even-simpler monad instantiation is
\lstinline{$~$p := OState MachineState},
where
\lstinline{OState(S A: Type) := S -> (option A) * S}
uses a \lstinline{None} answer to indicate that a failure occurred.
Its \lstinline{Bind} and \lstinline{Return} operations are implemented as
\begin{lstlisting}[basicstyle=\ttfamily\small]
Bind A B (m: OState S A) (f: A -> OState S B) :=
  fun (s: S) => match m s with (Some a, s') => f a s' | (None, s') => (None, s') end;
Return A (a: A) := fun (s: S) => (Some a, s)
\end{lstlisting}
and an unrecoverable (hard) failure can be implemented as
\begin{lstlisting}[basicstyle=\ttfamily\small]
fail_hard S A: OState S A := fun (s: S) => (None, s)
\end{lstlisting}
For compiler-correctness proofs, \lstinline{fail_hard} can be used to indicate that a situation occurred that the compiler is supposed to avoid, e.g. memory access at an invalid address, and a compiler-correctness proof then states that all valid source programs are translated to \riscv{} programs that never fail.

Moreover, if the compiler has been designed to emit code that does not use certain features, the \riscv{} specification can be simplified by implementing the primitives of \autoref{code:primitives} used by these features as just \lstinline{fail_hard}.
For instance, the compiler presented by \citet{lightbulb} emits code that does not depend on the CSRs, does not use floating-point operations or atomics, and assumes that there is no virtual memory and that the code always runs at the \lstinline{MachineMode} privilege level.
Therefore, the monad instantiation used to specify its correctness implements the primitives \lstinline{makeReservation}, \lstinline{checkReservation}, \lstinline{clearReservation}, \lstinline{getCSRField}, as well as \lstinline{unsafeSetCSRField}, \lstinline{getPrivMode}, \lstinline{setPrivMode}
of \autoref{code:primitives} as just \lstinline{fail_hard} (while the TLB- and floating-point-related methods were omitted altogether in the translation from Haskell to Coq).


\subsubsection{Nondeterminism}

\quad

\instance{nondeterministic execution}{choosing a monad that associates executions with mathematical sets of results (a possibility not available directly in Haskell)}

\hspace{3em}
One way to add nondeterminism is to use the nondeterministic option state monad,
\lstinline{OStateND S A := S -> option (A * S) -> Prop},
where the \lstinline{option}'s \lstinline{None} constructor is used to indicate failure, and \lstinline{option (A * S) -> Prop} can be thought of as the set of all possible outcomes.
Its \lstinline{Bind} and \lstinline{Return} operations are implemented as

\begin{lstlisting}[basicstyle=\ttfamily\small]
Bind A B (m: OStateND S A)(f : A -> OStateND S B) :=
  fun (s : S) (obs: option (B * S)) =>
    (m s None /\ obs = None) \/
    ($\exists$ a s', m s (Some (a, s')) /\ f a s' obs);
Return A (a : A) :=
  fun (s : S) (oas: option (A * S)) => oas = Some (a, s)
\end{lstlisting}

\subsubsection{Runtime input}

\quad

\instance{input and output}{combining nondeterminism with a state type extended to store a trace of interactions}

Once we have nondeterminism, we can use it to model memory-mapped I/O (MMIO).
For instance, in the implementation of the \lstinline{loadWord} primitive, if the address is not a physical memory address, we delegate to the following helper function:
\begin{lstlisting}[basicstyle=\ttfamily\small]
mmio_load32 addr: OStateND S int32 := fun s oas =>
  (isMMIOAddr addr /\ $\exists$ v: int32, oas =
    Some (v, (appendLog (mmioLoadEvent addr v) s))) \/
  (~isMMIOAddr addr /\ oas = None)
\end{lstlisting}
It can be read as a function that for each current state \lstinline{s} returns a proposition that indicates whether an outcome \lstinline{oas} (of type \lstinline{option (int32 * MachineState)}) is in the set of possible outcomes, distinguishing two cases based on whether the address lies in the address range reserved for MMIO.
We also augment \lstinline{MachineState} with a log to which we append an MMIO event on each load and store that falls into the MMIO address range.

Proof of a compiler targeting this specification will have to show that all states in the outcome set given by \lstinline{mmio_load32} satisfy the compiler's correctness guarantees (such as being related to a state of the source-language execution), so the body of \lstinline{mmio_load32} will appear on the left-hand side of an implication, so the existentially quantified \lstinline{v} becomes universally quantified, and as expected, the compiler has to prove that its guarantees hold for all possible values \lstinline{v} that this MMIO load could have read.

\subsubsection{Nondeterminism by means of weakest preconditions}

\quad

\instance{smooth integration with Hoare-logic-style program verification}{first assigning programs meanings in the style of interaction trees and then applying recursive functions (like weakest-precondition computation) to those trees}

The Bedrock2 compiler~\cite{lightbulb} using our \riscv{} specification requires \riscv{} semantics that given an initial state \lstinline{s}, a monadic computation \lstinline{m} corresponding to the execution of a sequence of primitives from \autoref{code:primitives}, and a desired \emph{postcondition}, returns the weakest precondition that must hold in order for the postcondition to hold.
Therefore, it seems that we need the following bridge definition that tells when a monadic \lstinline{OStateND} computation satisfies a postcondition:
\begin{lstlisting}[basicstyle=\ttfamily\small]
mcomp_sat S A (m: OStateND S A) s post :=
  $\forall$ o, m s o -> $\exists$ a s', o = Some (a, s') /\ post a s'
\end{lstlisting}
For an example relating this definition to the previous subsection, \lstinline{m} could be instantiated with \lstinline{mmio_load32 addr}, and \lstinline{post} could be instantiated with the claim that the final state is related to a state of the source-language execution.

When instantiating \lstinline{m} with a monadic computation involving many \lstinline{Bind}s, unfolding \lstinline{mcomp_sat} and all the \lstinline{Binds} quickly leads to huge formulas involving an existential for each intermediate state and answer, and we found these formulas to be larger than what human brains can deal with productively.
The solution was to treat \lstinline{mcomp_sat} and \lstinline{Bind} as opaque and to prove weakest-precondition-style rules for each primitive of \autoref{code:primitives}, using only these rules in the compiler-correctness proof, so that the large formulas were confined to just the proofs of these rules.

However, when a processor in the Coq-embedded hardware-description language Kami~\cite{Kami_ICFP17} was being proved against our \riscv{} specification, the same formula-explosion problem struck again, but this time, on the other (left-hand) side of the implication.
Inversion rules for \lstinline{mcomp_sat} of primitives, dual to the weakest-precondition-style rules mentioned above, might have been a way to go, but it turned out that it is simpler (both for the compiler and the processor) to use an instantiation that is more suitable for weakest-precondition generation, namely a free monad.

In the style of interaction trees~\cite{interaction_trees}, we use a Coq \lstinline{Inductive} for effects with one constructor per primitive of \autoref{code:primitives}, and a generic free monad with one constructor for an effect followed by a continuation, plus a second constructor to indicate termination.
\lstinline{Bind} for this monad can be defined as a \lstinline{Fixpoint} that flattens monadic computations that might have nested \lstinline{Bind}s as the first argument of \lstinline{Bind} into a more canonical form.
A result is almost a sequential list of effects (ended by the termination constructor of the free monad), except in the case of nondeterminism, where branching can occur, so the shape becomes a tree there.

On this free-monad structure, we can run an interpreter that computes weakest preconditions.
The crucial difference between \lstinline{OStateND} and the free-monad interpreter is that the former creates an existential for the intermediate state and answer of each \lstinline{Bind}, whereas the latter works similarly to a continuation-passing-style interpreter and just passes updated states to the right-hand sides of the \lstinline{Bind}s, leading to considerably simpler formulas.
For comparison, here is the helper function that the interpreter invokes in the \lstinline{loadWord} case when the address is not a physical memory address:
\begin{lstlisting}[basicstyle=\ttfamily\small]
mmio_load32 addr := fun s post =>
  isMMIOAddr addr /\ $\forall$ v: int32,
    post v (appendLog (mmioLoadEvent addr v) s)
\end{lstlisting}
Note how, contrary to \lstinline{OStateND}, no case for failure is needed, and the value \lstinline{v} being read is already universally quantified, rather than existentially quantified on the left-hand side of the implication of \lstinline{mcomp_sat}, and if more code follows after this snippet, it will be put into \lstinline{post} and thus be invoked with the updated state \lstinline{(appendLog (mmioLoadEvent addr v) s)}, with no intermediate existential.


\subsection{Multiplication in software: Reasoning about multiple instantiations of our spec}

\instance{proving connections between semantics variants (e.g. standing for capabilities of different conformant processors)}{simply instantiating the semantics with different type-class instances and mentioning the different instantiations in single theorem statements and proofs}

As a case study to showcase the benefits of instantiations of our spec being first-class objects, we present a Coq proof of the correctness of a trap handler that implements multiplication in software, for embedded processors that only implement the ``I'' extension of \riscv{} and thus raise exceptions when encountering multiplication instructions.

We want to show that a machine without hardware support for multiplication, but correctly configured with an exception handler that implements multiplication in software, behaves like a machine that supports multiplication in hardware.
This theorem could then be used to simplify reasoning about programs running on a machine without hardware multiplication, because it saves the burden of reasoning about the trap handler and instead makes it as easy as reasoning about the specification with multiplication in hardware:

\langCoq
\begin{lstlisting}[basicstyle=\ttfamily\small]
match inst with
| Mul rd rs1 rs2 => x <- getRegister rs1; y <- getRegister rs2; setRegister rd (mul x y)
| ...
end
\end{lstlisting}

Parts of the exception handler are implemented in the Bedrock2 source language~\cite{lightbulb} and compiled using the Bedrock2 compiler, but the handler also needs some low-level operations that are not expressible in the Bedrock2 source language and are therefore implemented by-hand in assembly.
Our proof combines a program-logic proof about the Bedrock2 handler function, the compiler-correctness proof, and a proof about the assembly instructions, guaranteeing that all these parts have been put together correctly, and the final statement only mentions \riscv{} semantics. All the other interfaces have been canceled out by combining the proofs and thus are not part of the trusted code base any more.

In this case study, three instances of our \riscv{} semantics are involved (where the first one does not appear in the final theorem statement but only inside its proof):
\begin{itemize}
\item An instance used by the compiler-correctness proof, which does not have any CSRs (control and status registers, required by the exception mechanism) in its state and fails (with undefined behavior) on all CSR-related instructions. For the compiler, this instance was chosen to simplify the proof, because the compiler does not emit any instructions that depend on CSRs.
\item The instance with hardware support for multiplication
\item The instance with a trap handler implementing multiplication in software
\end{itemize}

\subsubsection{The theorem statement}

We can state the theorem as follows:
\langCoq
\begin{lstlisting}[basicstyle=\ttfamily\small]
  Theorem softmul_correct: forall (initialH initialL: State) (post: State -> Prop),
    runsTo (mcomp_sat (run1 mdecode)) initialH post ->
    related initialH initialL ->
    runsTo (mcomp_sat (run1 idecode)) initialL (fun finalL =>
      exists finalH, related finalH finalL /\ post finalH).
\end{lstlisting}

It uses \lstinline{run1}, a function that defines how one single instruction is executed, which is parameterized over the instruction decoder, and to which we pass \lstinline{mdecode} (a decoder that supports the multiplication instruction) in the hypothesis and \lstinline{idecode} (a decoder that returns \lstinline{InvalidInstruction} for the multiplication instruction) in the conclusion:
\langCoq
\begin{lstlisting}[basicstyle=\ttfamily\small]
  Definition run1(decoder: Z -> Instruction): M unit :=
    pc <- getPC;
    inst <- Machine.loadWord Fetch pc;
    Execute.execute (decoder (LittleEndian.combine 4 inst));;
    endCycleNormal.
\end{lstlisting}

The \lstinline{mcomp_sat} function is of type \lstinline{M unit -> State -> (State -> Prop) -> Prop} and asserts that a monadic program (consisting of primitives of Figure
\ref{code:primitives}), applied to some initial state, satisfies a postcondition, and \lstinline{runsTo} lifts it to an arbitrary (but finite) number of steps.
The predicate \lstinline{related} is used to relate a high-level state (i.e. the state of a machine that supports multiplication in hardware) to a low-level state (i.e. the state of a machine that implements multiplication in software using a trap handler), and it also contains all the preconditions on how the low-level machine needs to be configured.
That is, \lstinline{related} asserts that the two states have the same values for the registers and the program counter, and that the memory (modeled as a partial map from 32-bit addresses to bytes) of the low-level machine contains everything of the high-level memory, as well as the instructions of the exception handler and some scratch space that the exception handler can use as its stack (which must be available even if the main program has used up all of its stack).
To define at which address in memory the handler and the scratch space are located, \riscv{} defines some control-and-status registers (CSRs) that our definition of \lstinline{related} mentions:
\begin{itemize}
  \item The CSR called \lstinline{MTVecBase} is used to store the address of the trap handler. 
  \item The CSR called \lstinline{MScratch} is a read/write register dedicated for use by machine mode, and we use it to store the address of the \emph{end} of this scratch space. 
\end{itemize}

So overall, the theorem \lstinline{softmul_correct} can be read as follows:
If a machine with hardware multiplication runs to a high-level state satisfying a postcondition, then every related machine with software multiplication runs to a low-level state which, when translated back to a high-level state, satisfies the same postcondition.


\subsubsection{The handler code}

The exception-handler code is implemented partially in handwritten assembly and partially in the Bedrock2~\cite{lightbulb} source language and compiled to bytes by the Bedrock2 compiler.
In order to \emph{prove} the \lstinline{softmul_correct} theorem, we use the correctness theorem of the Bedrock2 compiler, but note that the \emph{statement} of the \lstinline{softmul_correct} theorem does not depend on the Bedrock2 language semantics or on anything related to the fact that we used the Bedrock2 compiler, so the auditing burden for someone (who trusts the Coq proof checker) auditing our handler is much smaller, because one does not need to worry about the compiler, its language semantics, and its interaction with the assembly code.

The first few instructions of our handler (handwritten in Coq) are as follows:

\langCoq

\begin{minipage}{\linewidth}
\begin{lstlisting}[basicstyle=\ttfamily\small]
  Definition handler_init :=
  [[  Csrrw sp sp MScratch;    (* swap stack pointer (sp) and MScratch CSR *)
      Sw sp zero (-128);       (* save the 0 register (for uniformity) *)
      Sw sp ra (-124);         (* save ra *)
      Csrr ra MScratch;        (* use ra as a temporary register... *)
      Sw sp ra (-120);         (* ... to save the original sp *)
      Csrw sp MScratch;        (* restore the original value of MScratch *)
      Addi sp sp (-128)    ]]. (* remainder of code will be relative to updated sp *)
\end{lstlisting}
\end{minipage}


After that, the registers 3 to 31 are saved to the scratch space as well, and then the Bedrock2-generated part is called by passing it the value of the CSR register \lstinline{MTVal}, which contains the invalid instruction that caused the exception, and a pointer to the scratch space in which we saved the registers.
The Bedrock2 code is written directly in Coq using the custom-notations feature, a C-like syntax, and operator precedence as suggested by whitespace in this example:

\begin{lstlisting}[mathescape=false]
Definition softmul := func! (inst, a_regs) {
  a = a_regs + (inst>>15 & 31)<<2;
  b = a_regs + (inst>>20 & 31)<<2;
  d = a_regs + (inst>>07 & 31)<<2;
  unpack! c = rpmul(load(a), load(b));
  store(d, c)
}
\end{lstlisting}



It extracts the three 5-bit fields of the instruction that indicate the two source registers (operands of the multiplication operation) and the destination register, respectively, and then calls another Bedrock2 function \lstinline{rpmul} that implements multiplication in terms of addition, storing the result back into the scratch space.
The \lstinline{rpmul} function iterates over the bits of the second operand while repeatedly doubling the first operand, a technique sometimes called ``Russian peasant multiplication.''
Both \lstinline{softmul} and \lstinline{rpmul} are verified using the Bedrock2 program logic.
The spec of the former is:

\langCoq
\begin{lstlisting}[basicstyle=\ttfamily\small]
Instance spec_of_softmul : spec_of "softmul" :=
  fnspec! "softmul" inst a_regs / rd rs1 rs2 regvals R,
  { requires t m :=
      mdecode (word.unsigned inst) = MInstruction (Mul rd rs1 rs2) /\
      List.length regvals = 32 /\
      seps [a_regs |-> word_array regvals; R] m;
    ensures t' m' := t = t' /\
      seps [a_regs |-> word_array (List.upd regvals (Z.to_nat rd) (word.mul
               (List.nth (Z.to_nat rs1) regvals default)
               (List.nth (Z.to_nat rs2) regvals default))); R] m' }.
\end{lstlisting}

Its pre- and postcondition are expressed in terms of an (unused) I/O trace \lstinline{t} and the memory \lstinline{m}, for which we assert a list of two separation-logic clauses (a word array corresponding to the scratch space containing the register values, and a generic frame \lstinline{R} for the rest of the memory). 

By combining the program-logic proofs about the two Bedrock2 functions with the compiler-correctness theorem, we obtain that if we run the compiler within Coq to obtain a list of instructions \lstinline{mul_insts}, these instructions satisfy a verbose but unsurprising specification, laying out calling-convention details.

\subsubsection{Correctness proof of assembly part}

The assembly part of the handler is proven correct by induction over the \lstinline{runsTo} hypothesis of \lstinline{softmul_correct}.
If the machine with hardware multiplication executes any instruction besides multiplication, we just need to show that after executing the same instruction on the machine with software multiplication, the \lstinline{related} judgment is preserved, but we can do that once-and-for-all by inspecting each \emph{primitive} of Figure~\ref{code:primitives}, instead of analyzing the much larger number of \emph{instructions} that \riscv{} has.
The interesting case is when the machine with hardware multiplication encounters a multiplication instruction, and we have to show that the machine with software multiplication steps to a related state.
We do so by first symbolically executing the specification of what the \emph{hardware} does in case of an exception, which boils down to setting some CSR fields and then setting the PC to the exception-handler address found in the \lstinline{MTVecBase} CSR.
After that, we symbolically execute the handwritten assembly instructions, using Coq's proof context to keep track of all the facts that we know about the current state of the machine.
For each assembly instruction, we encounter its specification in terms of the primitives of Figure~\ref{code:primitives}, and for each primitive, we have a helper lemma that updates our symbolic state.
At the point where we reach the call to the Bedrock2-generated code, we apply the correctness lemma for the compiled trap handler.
After that call, we step through more handwritten assembly instructions that restore the registers and then call the \lstinline{Mret} instruction that jumps back to one instruction past the multiplication instruction that caused the exception.
At that point, we need to prove that the symbolic state accumulated in the Coq proof context implies that the two machines are still \lstinline{related}, which only works if there are no bugs in the handler code.

\subsubsection{Bugs found during verification}

At that final point in the proof described above, we actually found two interesting bugs.
The first one was that we forgot to reset the \lstinline{MScratch} CSR, so one invocation of the exception handler works fine, but the next one will use a wrong address for its scratch space.
The second bug was the corner case where the multiplication instruction stores its result into the stack pointer.
In that case, we must not override the stack pointer with the original stack pointer that we swapped into the \lstinline{MScratch} register at the beginning of the handler.

We also found two more obvious bugs related to when to set the stack pointer and what stack-pointer offsets to use.


\subsection{Model checking with weak memory models}\label{sec:weak_memory}

So far, in considering breadth of applications of our semantics, we have covered the classic examples of program interpretation and interactive proof.
We finish with another classic, model checking, first of software programs and then of hardware designs.

\instance{model-checking of all possible program executions under weak memory}{choosing a type-class instance that records information on alternative paths that should be tried later}

\langHaskell

In this section we outline our approach to instantiate the type class to generate all outcomes of small multicore litmus tests with respect to the memory model.
We instantiate the type class with a runtime implementing the exploration algorithm of \citet{HMC_ASPLOS20}.

This algorithm revolves around 4 data structures:
\begin{itemize}
\item a control/data/addr dependency-bookkeeping data structure, to maintain a list of all the memory events that imply dependencies on the currently interpreted instruction
\item  a current partial execution graph, which is the graph of the memory events and their memory-model relations
\item two bookkeeping data structures necessary for backtracking during search: a list of alternative partial execution graphs to explore later and a maintained set of all the read events that would be subject to revisiting, if a store to the same address would occur.
\end{itemize}

Intuitively, our implementation goes as follow: we write an interpreter in charge of exploring an execution path depth-first.
That interpreter also records all the alternative decisions that it could have taken on its way.
The interpreter can either return successfully with a valid execution, or it can return that the execution that it explored ended up violating the memory model.
In both cases, the interpreter updates the global bookkeeping of alternative executions.

More precisely, the interpreter is a classic interpreter except for the following twists.
\begin{itemize}
\item It keeps track of the dependencies that previous memory events have on the different registers. 

It is worth noting that this functionality does not need to be interleaved in the execution semantics.
Rather, it can be done directly from the decoded instruction without looking at register values.
The reason is that, for one instruction, dependencies are known statically.

\item On a load, the interpreter adds to the partial execution that the load reads from one of the stores to the same address already present in the current partial execution. The interpreter also adds all other possible alternative stores to the same address as alternative executions to explore later.
The interpreter then calls Alloy~\cite{alloy} with the partial execution graph that it just extended.
Alloy verifies that this extension is \riscv{}-compliant.
If not, the interpreter signals failure after updating backtracking structures; otherwise, it keeps running.
\item On a store, the interpreter both adds a new store event to the current execution and updates the alternative partial executions: any load in the revisit set that loads from the address of the current store is the source of a new partial execution to record.
\item When the interpreter reaches the end of one thread, it starts running the next thread.  When the interpreter finishes running the last thread successfully, it returns that the current execution is a valid execution.
\end{itemize}

This straightline interpreter does not do the backtracking itself.
Instead, it is called from a top-level loop that keeps calling the interpreter on the next partial execution to explore, each time either getting a valid execution or a failure but an updated backtracking structure. 


We implement this model-checking interpreter by instantiating the \verb|RiscvMachine| type class in the I/O monad. We use references to track state associated with the model-checking algorithm. We modify the \verb|Minimal64| machine described in Section~\ref{sec:simulation} and add the partial-execution graph and bookkeeping data structures described above. The \verb|RiscvMachine| instance implements dependency tracking in the \verb|loadWord|, \verb|storeWord|, and \verb|fence| primitives; other primitives are implemented similarly to the \verb|Minimal64| simulator.  Figure~\ref{code:mm_defs} shows excerpts from the definition of this instance.

\begin{figure} \begin{lstlisting}[language=Haskell,basicstyle=\ttfamily\small,columns=fullflexible,commentstyle=\color{gray},deletekeywords={Monad,Integral,Int,Bool,Maybe,fromIntegral},mathescape=false]
instance RiscvMachine (MaybeT (ReaderT Ptrs IO)) Int64 where
  loadWord Execute ad = do
      refs <- ask
      (ev, execution) <- readEnv refs
      if Set.member ev (domain execution)
        then getValFromDomain ev execution
        else do
          updateRevisitSet refs ev
          (writeToReadFrom, newExec, altExecs) <- generateReadExecs refs ev execution ad
          lift . lift $ writeIORef (r_currentExecution refs) newExec
          storeAltExecs refs altExecs
          checkAlloyModel writeToReadFrom newExec
    where ...
\end{lstlisting}
  \caption{Excerpts from the memory-model-checking instantiation of the \riscv{} type class}\label{code:mm_defs}
\end{figure}

The complete implementation is ~800 lines of Haskell.

We use the upstream official Alloy
specification~\cite{RVWMOGithub} for RVWMO, one of the several machine-readable forms available online for the \riscv{} weak memory model.

We only support word load and store instructions plus a TSO fence, as our intention was simply to demonstrate that one can implement state-of-the-art memory-model-exploration algorithms using our specification.
Hence we did not go through the implementation of the exploration for atomics and release/acquire fences, which, based on the study of \citet{HMC_ASPLOS20}, we predict would not require interestingly different ingredients.





\subsubsection{Running \riscv{} memory-model litmus tests}

To demonstrate the usefulness of our model-checker, we develop a simple test harness to symbolically execute litmus tests for the \riscv{} memory model. We use the litmus tests provided by \citet{RVLitmusGithub}, which were also used by the \riscv{} Memory Model Task Group during development of the ISA specification. These litmus tests were generated using the \texttt{diy} tool suite~\cite{diyManual} and test a range of possible memory-model behaviors including store buffering, load buffering, and message-passing; each test has several variants that add fences, address dependencies, control dependencies, or data dependencies to highlight possible behaviors of the \riscv{} memory model. Each test consists of some initialization conditions on registers or memory, the instructions to be run on each thread to test the behavior of interest, and a postcondition which is met if the behavior of interest occurred. For example, the message-passing litmus test has a postcondition testing for the case where the flag is set but the value was not set.

The architecture of our test harness is as follows. We first parse the litmus-test specification file into a simple symbolic representation. We then compile the initialization conditions and testing instructions into a small \riscv{} assembly file, which first initializes the register state of each thread and then executes each thread's instructions. We use \texttt{gcc} to compile the assembly file into an ELF file, which is then passed to our model-checking algorithm as described in \autoref{sec:weak_memory}. The model-checking algorithm is modified to store the final register state of each thread (upon reaching the stopping PC value). For every execution deemed valid by the Alloy solver, we check whether these final register states satisfy the test postcondition; if so, this is reported to the user.

It is important to note that the litmus-test postconditions do not constitute a specification for the memory model; rather, they indicate particular weak-memory-model behaviors that may or may not be allowed by the ISA or by implementations thereof. We therefore validate our model-checker by checking that it generates the same behaviors as the axiomatic \texttt{herd} and operational \texttt{rmem} model-checking tools~\cite{RVLitmusGithub}.


We are able to run all of the basic 2-thread litmus tests from the \riscv{} litmus-test suite, and we anticipate that adding support for more complex tests or higher numbers of threads would not pose significant engineering difficulties. On the basic 2-thread tests, wall-clock running times range from approximately 20 seconds for the smallest test cases to 3 minutes for the largest; we are able to run all 36 test cases in 50 minutes. Tests were run on a lightly loaded machine with a Haswell i7-5930K CPU and 64GB of DDR4 RAM running Arch Linux.
The performance bottleneck here is fully in interaction with Alloy, e.g. spawning a new Java Virtual Machine per query and no doubt tuning Alloy parameters suboptimally.



\subsection{Model-checking the decode and execute functions}

\instance{compilation to hardware circuits}{carefully tweaking interpreter-style instances to avoid unbounded types}

The riscv-formal project~\cite{riscvformalGithub} proposes a Verilog description of the Boolean function updating one cycle: assuming a single-cycle machine, it specifies how the register file and the memory are transformed by an arbitrary instruction from the backbone of the base ISA.
They also have infrastructure to model-check their Verilog description against other descriptions, for example the standard \riscv{} simulator Spike~\cite{spike}; and against real processors, using the Yosys tool.
As our final case study, we decided to connect our specification to that ecosystem, validating it against the existing riscv-formal specification.
It presumably is also possible to check our specification directly against processors using Yosys, though we have not carried out that experiment yet.

The tooling here works directly on Verilog code, so it was convenient to translate our specification into that language.
At some level, it forced us to confront challenges that are not fundamental, as a specification does not actually need to execute as a finite hardware circuit.
However, there could be other applications of compiling to hardware, like testing of processor designs on FPGAs, so we felt it was worth investing in this path.
We used the Haskell-to-hardware converter Clash~\cite{clash} to transform our specification, using a minimal state-monad instance, into a Verilog Boolean function, and the authors of riscv-formal model-checked that output against their reference riscv-formal to find discrepancies (as revealed by any single instructions of execution from matching initial states).

Interestingly enough, the Clash instance of the specification is quite similar to the \texttt{Minimal} instances.
However, the \texttt{Minimal} instances are not directly usable in Clash because they use a \texttt{Map} for the register file, and these potentially arbitrary-sized maps do not normalize well in Clash.
Instead, we use the \texttt{Vector} datatypes in Clash.
We also fought with instabilities of Clash's partial evaluation of programs, where sometimes it would fail to notice dead code or otherwise take advantage of predictability of some code spans.
As a result, we did make a few more lines of change to our spec just to save Clash some trouble.

\section{Related work}\label{sec:related}

Most recent work on multipurpose ISA specs has employed domain-specific languages toward ends similar to ours.
The \sail{}~\cite{SAIL_POPL19, SAIL_RISCV_Github} language is the highest-profile today for defining ISA semantics.
Work there can be helped by specialized features on a spectrum of generality, from examples like dependently typed bitvectors (fairly generic) to specialized support for instructions and their bit-level representations.
A DSL in this category requires new tools to translate into specification languages required across use cases, and languages in this tradition had not previously allowed application-agnostic semantics to be first-class objects in logics.

Another DSL specific to the ARM ISA family~\cite{armspec} received a lot of attention recently, thanks to systematic adoption for several of ARM's most important ISA variations.
That second DSL has also been translated automatically to Sail.
A notable predecessor to \sail{} was L3~\cite{l3}.
These DSLs have been used for encoding several significant mainstream ISAs beyond \riscv{}, and indeed it is possible that our approach would be less appropriate for legacy ISAs with complications and baggage beyond what we had to deal with in \riscv{}.

There has also been a good amount of past work writing ISA semantics directly in the languages of proof assistants.
For instance, \citet{FoxM10} defined and validated an ARM semantics in HOL4.
As in our semantics, theirs uses a monadic style for state-threading and incorporation of effects.
However, they implemented a single monad, rather than using parameterization over a monad as the central approach to applying one semantics in different use cases.
They also performed extensive validation across a few use cases, most notably in automated testing against processors.

Goel and Hunt~\cite{x86ACL2_VSTTE13, Goel_Thesis16} developed a detailed model of x86 in ACL2.
Like us, they observed that different use cases impose different constraints on the formalization:
On one hand, they want to execute their model efficiently to validate it against real x86 processors, and on the other hand, they want to prove correctness of programs against their x86 model.
To bridge between these different requirements, they use an abstract state representation for program verification and a concrete state representation for execution, define correspondence predicates between the two, and prove that all modifications preserve the correspondence. 
Our approach can be seen as generalizing theirs to consider more than two logic interpretations of a semantics.

Crafting semantics for multicore systems with weak memory models has become a substantial research area in its own right, with some of the earliest work on mainstream ISAs centered at the University of Cambridge~\cite{SarkarSNORBMA09,AlglaveFIMSSN09}.
With the \sail{} language and others, it had already been demonstrated that opcode meanings could be separated from memory-model definitions.
Our preliminary results show that the same should work with our semantics style.

ISA semantics is the natural meeting point of software and hardware proofs, though we have been surprised to see how few past multipurpose specifications have been used for substantial proofs on both sides.
Considering more single-purpose specs, CompCert~\cite{Leroy-Compcert-CACM} includes quite a few assembly-language backends with associated operational semantics (including for RISC-V), as far as we know not reused to reason about hardware.
The CakeML project has connected hardware and software proofs via a semantics for the Silver ISA~\cite{CakeMLSilver_PLDI19}, where, to our knowledge, neither that ISA nor its semantics have been applied outside that case study.
The other two significant examples of that kind (around bespoke, verification-motivated ISAs) came earlier, in the CLI stack~\cite{bevier1989an} and the VeriSoft project~\cite{Verisoft}.

The most-involved functional-correctness proofs we know of, connected to prior multipurpose specifications of industrial ISAs, relate to \sail{} ARM specs, proving correctness of address translation in Isabelle/HOL~\cite{SAIL_POPL19} (with an ongoing port to Coq~\cite{SailCoqUseGithub}) and then capability security of the Morello ISA extension~\cite{Morello}.
These results are metatheorems about the semantics themselves, which require a different sort of detail-oriented reasoning than proof of particular significant machine-code programs or of processors.
In contrast, our specification has been validated through a case study~\cite{lightbulb} doing a complete functional-correctness proof for a simple embedded system, connecting proofs of hardware and software parts into a final theorem whose statement does not depend on our semantics.
The software and hardware sides of that project are compatible with mainstream \riscv{} artifacts: the verified software runs on an off-the-shelf \riscv{} microcontroller, while the verified processor also runs \riscv{} machine code produced by GCC.

Sail ISA semantics have also recently been connected to program proof with Islaris~\cite{Islaris}, a framework that uses SMT-based symbolic execution of the semantics to produce more manageable verification conditions, to be discharged in Coq.
The verification experience seems to be quite streamlined, though, contra our approach, at least one new trusted translator between languages is involved.
The SMT solver is also trusted, rather than somehow using it to generate proofs checkable by Coq.
Islaris has been applied successfully to both \riscv{} and more complex ARM ISAs.

\section{Conclusion}\label{sec:conclusion}

We have presented a new approach to formal specification of hardware instruction sets, relying on type classes for easy instantiation to different use cases, thus avoiding any requirement for domain-specific language features.
As a result, such a semantics can be written directly in popular general-purpose languages and translated to other forms using tools that are not domain-specific.
Our example is for the up-and-coming \riscv{} ISA family and has been applied across hardware and software and across different styles of formal methods, without requiring that a single new parser/translator be written to integrate with tools backed by interactive proof (Coq), relational model finding (Alloy), and SMT-based model checking (Yosys).

Some rough edges certainly remain, which should be considered in comparing our approach to that of domain-specific languages, motivating the search for new coding patterns to achieve better modularity in general-purpose languages.
We would rather not have our \lstinline{RiscvMachine} type class always contain the \lstinline{getFPRegister} and \lstinline{setFPRegister} primitives.
It would be better if they were only present if the floating-point extension is supported, and similarly for \lstinline{makeReservation}, \lstinline{checkReservation}, and \lstinline{clearReservation} required by the atomics extension.
The instruction-decode function is quite large, and every property about it that we proved (or tried to prove) in Coq leads to performance issues that require very careful, performance-aware proof engineering.



Nonetheless, this approach allowed for a small arsenal of nontrivial tools to be constructed fairly quickly, and related mechanizations may have a place in future formal-methods ecosystems.

\bibliographystyle{ACM-Reference-Format}
\bibliography{riscv-semantics}


\end{document}